\documentclass[traditabstract,twocolumn]{aa}

\usepackage{epsfig,graphicx,natbib}
\usepackage{longtable}
\usepackage{amssymb}
\usepackage{amsmath}
\usepackage{color}
\usepackage{subcaption}

\usepackage{todonotes}

\begin{document}

\title{Polarization calibration techniques for new-generation VLBI}

\author{I. Mart\'i-Vidal \inst{1,2}
  \and A.~Mus \inst{1,2}
  \and M.~Janssen \inst{3,4}
  \and P.~de~Vicente\inst{5}
  \and J.~Gonz\'alez\inst{5}
}

\institute{
  Departament d’Astronomia i Astrof\'isica, Universitat de Val\`encia, C. Dr. Moliner 50, 46100 Burjassot ,Val\`encia, Spain
  \and
  Observatori Astron\`omic, Universitat de Val\`encia, Parc Cient\'ific, C. Catedr\`atico Jos\'e Beltr\'an 2, 46980 Paterna, Val\`encia, Spain
  \and
  Max-Planck-Institut f\"ur Radioastronomie, Auf dem H\"ugel 69, D-53121 Bonn, Germany 
  \and
  Department of Astrophysics, Institute for Mathematics, Astrophysics and Particle Physics (IMAPP), Radboud University, P.O. Box 9010, 6500 GL Nijmegen, The Netherlands 
  \and
  Centro Astron\'omico de Yebes (IGN), Apartado 148, E-19180 E-Yebes, Spain
}

\date {Received  / Accepted}

\titlerunning{Polarization calibration for new-generation VLBI}
\authorrunning{Mart\'i-Vidal et al.}

\abstract{The calibration and analysis of polarization observations in Very Long Baseline Interferometry (VLBI) requires the use of specific algorithms that suffer from several limitations, closely related to assumptions in the data properties that may not hold in observations taken with new-generation VLBI equipment. Nowadays, the instantaneous bandwidth achievable with VLBI backends can be as high as several GHz, covering several radio bands simultaneously. In addition, the sensitivity of VLBI observations with state-of-the-art equipment may reach dynamic ranges of tens of thousands, both in total intensity and in polarization. In this paper, we discuss the impact of the limitations of common VLBI polarimetry algorithms on narrow-field observations taken with modern VLBI arrays (from the VLBI Global Observing System, VGOS, to the Event Horizon Telescope, EHT) and present new software that overcomes these limitations. In particular, our software is able to perform a simultaneous fit of multiple calibrator sources, include non-linear terms in the model of the instrumental polarization and use a self-calibration approach for the estimate of the polarization leakage in the antenna receivers. } 

\maketitle

\section{Introduction}

The technique of Very Long Baseline Interferometry (VLBI) provides the highest angular resolutions in the study of the Universe. With this technique, it is possible to synthesize an aperture similar to the Earth diameter by combining the signals from radio telescopes distributed across the planet surface \citep[e.g.][]{TMS}. VLBI is based on the Van Cittert-Zernike theorem, a direct Fourier relationship between the sky brightness distribution and the spatial coherence of the electromagnetic field that reaches the Earth.

The heterodyne receivers of the VLBI radio telescopes are able to split the sky signal into two orthogonal polarization channels (usually, right-handed and left-handed circular polarizations; RCP and LCP, respectively), which are recorded independently. This splitting allows us to recover the full-polarization information of the observed sources by extending the van Cittert-Zernike theorem to the four Stokes parameters, via the so-called Radio Interferometer Measurement Equation \citep[RIME;][]{Hamaker}.

The RIME relates the complex visibilities (i.e., a measurement of the spatial coherence of the radio waves), computed between the polarization channels of two radio telescopes, to the Fourier transforms of the four Stokes parameters of the observed source. The RIME also accounts for the different instrumental effects (from electronic gains to atmospheric effects and different kinds of instrumental polarization).

The problem of the correct calibration of the instrumental polarization in astronomical interferometry has already been studied for a long time \citep[e.g.,][]{Cotton93,Leppanen95,IMVGMVA}. However, the most used polarimetry algorithms to date suffer from several limitations based on several assumptions that, even though well justified for the mainstream radioastronomical hardware that was used a decade ago, may not be suitable for the new-generation instrumentation. Assumptions like a constant instrumental polarization across the frequency bandwidth of a receiver, or the use of a linearized model to describe the leakage between antenna polarizers, are too simplistic for the calibration of (ultra-)wideband feeds and high dynamic-range observations.

In this paper, we discuss the main limitations of the classical algorithms used for the calibration of the instrumental polarization in VLBI and present new software, based on the RIME, that overcomes such limitations in different ways. In the next section, we introduce the basic aspects of VLBI polarimetry. In Sect. \ref{LPCALDescSec}, we describe the polarimetry algorithm used by the majority of the VLBI community and emphasize its limitations. In Sects. \ref{PSolveDescSec} and \ref{SimulSec}, we describe our software and show realistic simulations of VLBI observations, 
where the advantages of our algorithm can be quantitatively assessed. In Sects. \ref{sec:GMVA} and \ref{sec:VLBA}, we apply our algorithm to real Global mm-VLBI Array (GMVA) and Very Long Baseline Array (VLBA) full-polarization observations of sources 3C\,345 and 3C\,279, previously reported in \cite{IMVGMVA} (hereafter MV12) and \cite{KravVLBA} (hereafter K20). In Sect. \ref{sec:summary}, we summarize our conclusions.

\section{VLBI Polarimetry Modelling}

In the RIME framework \citep[][]{Hamaker,Smirnov}, the complex visibilities between two VLBI antennas, $a$ and $b$, are formulated as a visibility matrix, $\mathbf{V}^{ab}$, of the form

\begin{equation}
\mathbf{V}^{ab} = \begin{bmatrix} V^{ab}_{rr} & V^{ab}_{rl} \\ V^{ab}_{lr} & V^{ab}_{ll} \end{bmatrix},
\label{VMatrixEq}
\end{equation}

\noindent where $r$ and $l$ refer to the RCP and LCP polarization channels. The measured visibilities are a combination of two independent contributions: on the one hand, the instrumental plus atmospheric effects and on the other hand the brightness distribution of the four Stokes parameters of the observed source. Regarding the instrumental contribution to the signal polarization, there are several quantities involved,
from the relative gains between polarizers (i.e., differences in phases, delays and amplitudes between RCP and LCP), to the signal leakage between polarizers (i.e., the amount of RCP signal that is being injected into the LCP channel, and 
vice versa, with a given amplitude and delay/phase). All these instrumental and atmospheric quantities can be arranged in the form of one single Jones matrix. For antenna $a$, this matrix is equal to \citep{Smirnov}

\begin{equation}
\mathbf{J}^a =\mathbf{G}^a \mathbf{D}^a \mathbf{P}^a =  \begin{bmatrix} G^a_r & 0 \\ 0 & G^a_l \end{bmatrix} \begin{bmatrix} 1 & D^a_r \\ D^a_l & 1 \end{bmatrix} \begin{bmatrix} e^{j\phi_a} & 0 \\ 0 & e^{-j\phi_a} \end{bmatrix},
\label{JonesEq}
\end{equation}

\noindent where $j$ is the imaginary unit, the quantities $D^a_r$ and $D^a_l$ (called ``D-terms'') are used to model the signal leakage between the polarization channels (the D-terms may depend on frequency, but are assumed to be stable in time during the extent of the observations); $G^a_r$ and $G^a_l$ account for the electronic gains (plus the atmospheric delay/opacity) at each polarizer, and $\phi_a$ is the angle of the feed orientation of antenna $a$ when pointing to the observed source. $\mathbf{P}^a$ is the matrix that accounts for the effect of the feed angle in the source frame, $\mathbf{G}^a$ is the ``Gain matrix'', and $\mathbf{D}^a$ is the ``D-term matrix''.

Regarding the contribution of the observed source to the RIME, the equation relates the complex visibilities between antennas $a$ and $b$ to the so-called ``Brightness matrix'', $\mathbf{B}^{ab}$, which contains the brightness distributions of the four Stokes parameters (namely, $\mathcal{I}$, $\mathcal{Q}$, $\mathcal{U}$, and $\mathcal{V}$) in the Fourier domain, via the matrix equation

\begin{equation}
 (\mathbf{J}^a)^{-1} \mathbf{V}^{ab} ((\mathbf{J}^b)^H)^{-1} = \mathbf{B}^{ab} = 
 \begin{bmatrix} \tilde{\mathcal{I}}+\tilde{\mathcal{V}} & \tilde{\mathcal{Q}}+j\tilde{\mathcal{U}}  \\  \tilde{\mathcal{Q}}-j\tilde{\mathcal{U}}  &  \tilde{\mathcal{I}}-\tilde{\mathcal{V}} \end{bmatrix}.
\label{RIMEEq}
\end{equation}

The Fourier transforms ($\tilde{\mathcal{I}}$, $\tilde{\mathcal{Q}}$, $\tilde{\mathcal{U}}$, $\tilde{\mathcal{V}}$) of the Stokes parameters are evaluated at the point in Fourier space given by the baseline between antennas $a$ and $b$, projected onto the plane orthogonal to the source direction \citep{TMS}. All the instrumental effects are encoded in the Jones matrices, $\mathbf{J}^a$ and $(\mathbf{J}^b)^H$, where $H$ is the Hermitian operator. We assume that direction-dependent instrumental effects are negligible, which holds for the small fields of view achievable with VLBI.

If the polarized structure of the observed source cannot be resolved by the synthesized resolution of the interferometer, the solution to Eq. \ref{RIMEEq} is relatively simple, given that the Fourier transforms of all Stokes parameters are constant at all the points in Fourier space being sampled by the instrument. This special case allows us to model the source contribution to Eq. \ref{RIMEEq} using one real number per Stokes parameter, and the calibration problem becomes trivial. However, the high spatial resolutions achieved in VLBI, especially at high frequencies, imply that almost all the observable sources (even the most distant and compact calibrators) have some resolvable structures with non-constant (and complex-valued) Fourier transforms. 

Designing an accurate way to model the structure contribution of a spatially-resolved polarization calibrator in Eq. \ref{RIMEEq} is the key to any VLBI polarization calibration algorithm. The different approaches to model the contribution from the source polarization can be divided into three main families: (1) simple parametrization of the polarization source structure \citep[e.g., describing the brightness distributions using parametric models, as done in ][]{UVMultiFit}. (2) use of (non-parametric) deconvolution techniques in full polarization with an iterative fitting of the D-terms, as is done in the new versions of the EHT imaging library \citep{EHTimRef}. Another approach to use full-polarization deconvolution algorithms in the D-term fitting is to reconstruct the source polarization iteratively, using the incrementally-calibrated visibilities, and then fix the source
modeling to estimate the instrumental polarization in a forthcoming iteration. This strategy was first proposed by \cite{Cotton93}.

The third family of algorithms refer to a complete exploration of the whole parameter space using, for instance, Markov Chains (MCMC), as is done in the new versions of the EHT Themis library \citep{ThemisRef} and DMC \citep{dom2020}. These MCMC methods are able to recover the posterior distributions of the D-terms and their impact on each part of the reconstructed images, but their main disadvantage is the computational costs, especially for cases where the dimension of the parameter space makes any Monte Carlo exploration impractical (for example, the case of multi-calibrator geodesy observations, as discussed in Sect. \ref{WideFieldSec}).

\subsection{LPCAL}
\label{LPCALDescSec}

The \texttt{LPCAL} program \citep{Leppanen95}, provided as a
task in the Astronomical Image Processing System (AIPS) of the National Radio Astronomy Observatory (NRAO)\footnote{\texttt{http://www.aips.nrao.edu}.} is the best known polarimetry algorithm and that most used by the VLBI community. It has been tested and exploited for decades, to the point that the procedures followed by this algorithm can be considered as the standard for VLBI interferometric polarimetry. \texttt{LPCAL} models the source polarization using a parametric approach. The total-intensity brightness distribution, $\mathcal{I}$, is divided in a set of disjoint 
regions (or ``polarization sub-components''), $\mathcal{I}_i$, such that

\begin{equation}
\mathcal{I} = \sum_i^N{\mathcal{I}_i}.
\label{PSolveIEq}
\end{equation}

Each of these sub-components is assumed to have a constant fractional polarization. Hence, the brightness distributions of $Q$ and $U$ will be

\begin{equation}
\mathcal{Q} = \sum_i^N{q_i \mathcal{I}_i} ~~;~~ \mathcal{U} = \sum_i^N{u_i \mathcal{I}_i},
\label{PSolveQUEq}
\end{equation}

\noindent where all $q_i$ and $u_i$ are real-valued parameters defined in the interval $ q_i^2 + u_i^2 \in [0,1]$. We notice, though, that the \texttt{LPCAL} algorithm imposes no bounds to these parameters. Regarding $\mathcal{V}$, it is given by the difference between the calibrated parallel-hand correlations (i.e., the diagonal elements of the brightness matrix, Eq. \ref{RIMEEq}) whereas the Dterms may dominate the cross-hand correlations. Therefore, $\mathcal{V}$ can be neglected in the calibration of the Dterms.

The \texttt{LPCAL} approach allows us to model relatively complex brightness distributions using a minimum set of parameters for the source polarization. In the \texttt{LPCAL} procedure, the complex source structure in total intensity is first deconvolved using, e.~g., the CLEAN algorithm \citep{Hogbom}, and such a structure is then divided into regions of assumed constant polarization (i.e., the algorithm assumes that the fractional polarization and EVPA of the source is constant across the extent of each polarization sub-component; $\mathcal{I}_i$ in Eq. \ref{PSolveIEq}). This is an enhanced version of the so-called ``similarity assumption'' \citep{Cotton93}. Depending on how the source division is done, the \texttt{LPCAL} source model may deviate from reality in cases of sources with complex polarization structures \citep[for instance, sources with a strongly polarized emission that may be shifted with respect to the peak of $\mathcal{I}$, as regularly happens in AGN jets; e.g.,][]{Sullivan11,IMVGMVA, AgudoGMVA,GomezRef}.

The \texttt{LPCAL} algorithm estimates the parameters that define the source polarization (i.e., $q_i$ and $u_i$ in Eqs. \ref{PSolveQUEq}) and the D-terms of all antennas simultaneously, by means of a least-squares minimization of the error function given by \citep{Leppanen95}

\begin{equation}
\chi^2 = \chi^2_{rl} + \chi^2_{lr},
\label{Chi2LPCALEq}
\end{equation} 

\noindent with

\begin{equation}
\chi^2_{rl} = \sum_{k}{ \left| \frac{V^k_{rl}}{G^a_r (G^b)^*_l} - \frac{D^a_r V^k_{ll}}{G^a_l (G^b)^*_l} - \frac{(D^b)^*_l V^k_{rr}}{G^a_r (G^b)^*_r} - \mathcal{P}_{1,k}  \right|^2 w_k}
\label{RLLPCALEq}
\end{equation} 

\noindent and 

\begin{equation}
\chi^2_{lr} = \sum_{k}{\left| \frac{V^k_{lr}}{G^a_l (G^b)^*_r} - \frac{D^a_l V^k_{rr}}{G^a_r (G^b)^*_r} - \frac{(D^b)^*_r V^k_{ll}}{G^a_l (G^b)^*_l} - \mathcal{P}_{2,k}  \right|^2 w_k}.
\label{LRLPCALEq}
\end{equation}

\noindent where $\omega_k$ is the weight of the $k$-th visibility.
In these equations, $\mathcal{P}_{1,k} = (\tilde{\mathcal{Q}}_k+j\tilde{\mathcal{U}}_k)e^{j(\phi_{a}+\phi_{b})}$ and $\mathcal{P}_{2,k} = (\tilde{\mathcal{Q}}_k-j\tilde{\mathcal{U}}_k)e^{-j(\phi_{a}+\phi_{b})}$, where the Fourier transforms of the Stokes parameters, $\tilde{\mathcal{Q}}_k$ and  $\tilde{\mathcal{U}}_k$, are evaluated at the point of uv-space corresponding to the $k$-th visibility. Notice that the antenna indices, $a$ and $b$, depend implicitly on the visibility index, $k$, as well as the complex gains, $G^a$ and $G^b$ (for both polarizers, $r$ and $l$), which must be known for all the observing times of the experiment. The antenna gains can be estimated by means of self-calibration to Stokes $\mathcal{I}$ \citep{selfcalRef}. In these equations, we have omitted $k$ in the gains and the antenna indices, for clarity. 

Thanks to the parallactic-angle coverage in a set of VLBI observations (i.e., the time changes in $\phi_a$ and $\phi_b$) it is possible to decouple the D-terms from the source polarization parameters in Eqs. \ref{RLLPCALEq} and \ref{LRLPCALEq}. Essentially, the more complete the $\phi$ coverage in a VLBI experiment, the more accurate the D-term calibration.

\subsubsection{Limitations of LPCAL}
\label{LimitsSec}

Equations \ref{Chi2LPCALEq}, \ref{RLLPCALEq} and \ref{LRLPCALEq} are an approximation to the exact equation

\begin{equation}
\chi^2_{\mathrm{RIME}} =  \sum_{k,m}{w_k\left( \left| \left[ (\mathbf{J}^a)^{-1} \mathbf{V}^{ab} ((\mathbf{J^b})^{-1})^H \right]_{m} - \mathbf{B}_m \right|^2 \right)},
\label{Chi2PsolveEq}
\end{equation} 

\noindent where $m$ runs over the $rl$ and $lr$ matrix elements of Eq. \ref{RIMEEq}. This exact RIME version of the polarimetry error function contains several elements that are proportional to products among the different antenna D-terms. However, the \texttt{LPCAL} model (Eq. \ref{Chi2LPCALEq}) only contains elements that depend linearly on the D-terms. As long as the D-terms (the amount of signal leakage between antenna polarizers) are small enough and the signal is limited by sensitivity (i.e., the dynamic range of the polarization is low), the linear approximation of \texttt{LPCAL} should be good enough for the polarization calibration. However, modern VLBI interferometers may need the use of the exact RIME equation, as we discuss in Sect. \ref{SimulNLSec}.

The \texttt{LPCAL} algorithm also suffers from other limitations that, in special situations, can be critical. For example, the D-terms of all the antennas appearing in the data always have to be solved, with no restrictions nor use of any a-priori information about their polarization leakage. If, for some reason, the D-terms of a given antenna are known (e.g., they are provided by the corresponding observatory), that information would not be usable by \texttt{LPCAL}. Either all the data related to that antenna would not be used by \texttt{LPCAL}, or the D-terms of that antenna would have to be fitted, together with those of the remaining antennas, with no restrictions on their values. Furthermore, if there is an antenna with a missing polarization channel (i.e., only the data of one circular polarizer have been properly recorded), \texttt{LPCAL} cannot estimate the D-terms for that antenna. The reason for this limitation can be seen in Eqs. \ref{RLLPCALEq} and \ref{LRLPCALEq}, where the observed parallel-hand visibilities of both polarizations, $V_{rr}$ and $V_{ll}$, are explicitly needed for the construction of the model of both, $V_{rl}$ and $V_{lr}$.

In addition to these issues, \texttt{LPCAL} can only use one calibrator at a time, which may have important limitations in some VLBI experiments where some of the antennas may have a very limited parallactic-angle coverage for the source being used as calibrator. Last but not least, there is no way to parameterize any frequency dependence of either the instrumental or the source-intrinsic polarization, which is a clear limiting factor for the analysis of wide-band VLBI observations.

\subsection{PolSolve}
\label{PSolveDescSec}

We have developed a new code, \texttt{polsolve}, that overcomes the limitations of \texttt{LPCAL} described in the previous section. Our code is written in C++ and compiled as a Python module for use in the NRAO's Common Astronomy Software Applications (CASA) software infrastructure\footnote{\texttt{https://casa.nrao.edu}}, to which VLBI capabilities have recently been added \citep{vanBemmel2018, Janssen2019}. Our algorithm is part of a VLBI polarimetry toolbox that we have developed for CASA and is publicly available with a GPLv3 license\footnote{\texttt{https://code.launchpad.net/casa-poltools}}. The \texttt{polsolve} algorithm estimates the D-terms and the source polarization parameters from the least-squares minimization of Eq. \ref{Chi2PsolveEq}, which is the error function computed with the exact RIME. Hence, \texttt{polsolve} is not limited to the case of small polarization leakage at the antennas and/or low fractional source polarizations. Our algorithm is also able to use many calibrator sources (each one with its own polarization brightness modelling) in a simultaneous fit, which may help optimizing the parallactic-angle coverage limitations in a VLBI array.

Regarding the instrumental and source-intrinsic frequency dependence of wideband observations, \texttt{polsolve} offers different possibilities. On the one hand, the D-terms can be expanded as a Taylor polynomial (of any degree) in frequency space. On the other hand, we have implemented the so-called ``multi-IF'' fitting mode, where the D-terms are allowed to take different values at each frequency channel, but the source polarization is forced to be consistent across the full bandwidth (by consistent, we mean that the source polarization is modelled using the same parameter values for all subbands, with the option of adding frequency de-polarization and/or Faraday rotation; see Sect. \ref{WideFieldSec} for an example). The advantage of the multi-IF approach is that any frequency dependence of the D-terms that cannot be parameterized with a smooth polynomial of low degree (e.g., oscillatory trends, jumps close to the baseband boundaries, etc.) can still be properly described with a minimum number of fitting parameters. A detailed example of the multi-IF fitting is given in Sect. \ref{SimulSec}.  

In the case of antennas with missing polarization channels, \texttt{polsolve} uses the model prediction of the parallel-hand visibilities (obtained from the source deconvolution), so that the polarization leakage of the single observed polarization channel can still be determined.

The structure of the source polarization can be parameterized in the same way as \texttt{LPCAL} (i.e., by partitioning the structure as in Eqs. \ref{PSolveIEq} and \ref{PSolveQUEq}, neglecting $\mathcal{V}$), but \texttt{polsolve} also gives the possibility of using an iterative approach, where a full-polarization source model (obtained via, e.g., CLEAN, executed on all Stokes parameters) can be fixed in the fit of the instrumental polarization. This strategy (which can be called ``polarimetry self-calibration'') was initially proposed by \cite{Cotton93}, although its use has been very limited in the VLBI community, likely due to the fact that it is not directly implemented in the well-known (and most used) \texttt{LPCAL} program. Actually, a more recent program called \texttt{GPCAL} \citep{Jongho}, based on \texttt{LPCAL}, has finally implemented this long-sought feature in AIPS. The polarimetry self-calibration iterations can be easily implemented in \texttt{polsolve} via the use of helper functions that are provided with the code. In Sect. \ref{PolSelfCal}, we present a case where the performance of the \texttt{LPCAL} source parametrization is compared to the polarimetry self-calibration.

\section{Synthetic Data Tests}
\label{SimulSec}

Each of the limitations described in Sect. \ref{LimitsSec} may have an impact on observations performed with modern VLBI systems. The wide fractional bandwidths introduce chromatic effects in the instrumental polarization, given that the systems are designed to have an optimum polarization response at a given nominal frequency in each band; departures from that frequency will thus  introduce unavoidable leakage effects. On the other hand, the high sensitivity of the systems may make the non-linear D-term effects detectable in the data.

We have performed a quantitative study of the effects that the limitations in \texttt{LPCAL} may have on the calibration of current VLBI observations. For this study, we have simulated several sets of VLBI observations using our CASA polarization toolbox. We have studied the effects of wide fractional bandwidths and limited parallactic-angle coverage by generating synthetic VLBI Global Observing System (VGOS) observations performed in geodetic mode. The non-linear effects of the instrumental polarization and the limitations related to the \texttt{LPCAL} source parametrization have been studied by simulating a VLBI experiment at 230\,GHz with the Event Horizon Telescope (EHT) array. In this case, we have used high leakage factors and a simulated source with a high fractional polarization. In the next subsections, we describe these simulations in more detail. In the package provided with the \texttt{polsolve} code, we have included all the simulation and analysis scripts needed to reproduce the results shown in the following subsections.

\subsection{Ultra-Wide Frequency Band: VGOS Simulations}
\label{WideFieldSec}

VGOS is a World-wide new-generation interferometer, developed by the International VLBI Service (IVS), aimed at high-precision (up to sub-millimeter) geodesy observations. To achieve the SNR required for such an accuracy, the VGOS antennas are equipped with ultra-wide receivers, covering from 2 to 15\,GHz \citep[e.g][]{VGOSREF}. The VGOS bandwidth makes it necessary to observe in a basis of linear polarization. Such a polarization basis prevents a proper phase calibration of the resulting VLBI observations, unless the visibility matrices are converted to a circular (i.e., RCP and LCP) basis \citep{POLCONVERTREF}. The \texttt{polconvert} algorithm has already been successfully applied for the conversion of VGOS data from linear to circular basis \citep{AlefEUVGOS}. 

The estimates of \texttt{polconvert} for the cross-polarization antenna gains are based on the ``cross-polarization global fringe fitting'' technique \citep[equation 16 of][]{POLCONVERTREF}, which has limitations for cases of poor parallactic-angle coverage and/or calibrator observations with a low signal-to-noise ratio. As a consequence, any small deviation in the estimate of the phase difference, $\Delta \alpha_a$, between the linear polarizers of any antenna, $a$, and the \texttt{polconvert} estimate will be converted into a polarization leakage term in the circular-basis visibility matrices \citep{APPRef,CiriacoRef}. It can be shown \citep{CiriacoRef} that the D-terms corresponding to the polarization conversion with a $\Delta \alpha_a$ phase offset are pure imaginary and equal for RCP and LCP. Such a leakage corruption in the VGOS polarimetry can be corrected with the use of \texttt{polsolve}, as we discuss in this section.

We have used the \texttt{polsimulate} CASA task (which is part of our CASA toolbox for VLBI polarimetry) to simulate a set of VGOS observations with a large fraction of the C band frequency coverage of the system (from 4\,GHz to 6\,GHz). Our frequency setup consists of 64 spectral windows distributed homogeneously across the simulated bandwidth. We have introduced a chromatic (frequency-dependent) phase offset between the polarizers of all the antennas, to simulate the residual post-conversion leakage after a sub-optimal run of \texttt{polconvert}. The simulated frequency dependence of the phase offsets is sinusoidal, with an amplitude of 10 degrees \citep[which would introduce a maximum polarization leakage of the order of 10\%; ][]{CiriacoRef} and random phases and periods. The simulated VGOS array consists of the triplet of European sites currently operative \citep{AlefEUVGOS}. These are located at the Onsala Space Observatory (Sweden), the Wettzel Observatory (Germany) and the Yebes Observatory (Spain). 

The main goal of the VGOS Project is to perform a quasi-continuous monitoring of VLBI geodetic parameters, via the observation of ICRF3-related AGN \citep[about 300 defining sources and more than 4500 supplementary sources;][]{ICRFREF}. Therefore, our simulations consist of a simplified version of a geodesy-like experiment, with observations of ten ICRF3 sources, organized in snapshot scans (30 seconds long) distributed along a total of 2 hours. Each source is only observed in two scans during the experiment, with a typical time separation of half an hour. It is clear that the parallactic-angle coverage of each source, which is what allows to perform a robust calibration of the D-terms, is minimal in this simulation. 

The simulated source structures consist of polarized point sources with random fractional polarizations, $p/I$, (with $p/I$ ranging between 1\% and 10\%), random electric-vector position angles, $\chi$, random spectral indices, $\alpha$ (with absolute values between 0 and 1), and random
rotation measures, $RM$, introduced by external Faraday screens (with $RM$ ranging, in absolute value, between 0 and 100\,rad/m$^2$). The flux densities of all sources are fixed to 1\,Jy (at the fiducial frequency of 4\,GHz). The simulated properties of the VGOS sources are given in Table \ref{VGOSTab}.
The system temperatures of the antennas are set to 50\,K across the whole bandwidth, which results in an SNR of $\sim 0.6-6$ for the cross-hand correlations, $V_{rl}$ and $V_{lr}$, at each spectral window and scan.

\begin{table}
\centering
\begin{tabular}{ c r r r r }
Source & \multicolumn{1}{c}{Q/I ($\nu_0$)} & \multicolumn{1}{c}{U/I ($\nu_0$)} & \multicolumn{1}{c}{RM} & \multicolumn{1}{c}{$\alpha$} \\
       &               &               &  \multicolumn{1}{c}{(rad/m$^2$)}   &            \\
\hline
 0320$+$415 & $0.0356$  & $-0.0254$ & $19.7$ & $0.46$  \\ 
 0716$+$714 & $-0.0091$ & $0.0222$  & $73.2$ & $-0.89$ \\
 0836$+$710 & $-0.0554$ & $0.0322$  & $93.9$ & $-0.96$ \\
 0010$+$405 & $-0.0756$ & $0.0387$  & $-63.3$ & $-0.64$ \\
 0112$-$017 & $0.0356$  & $0.0114$  & $-41.8$ & $-0.14$ \\
 1954$+$513 & $-0.0118$ & $0.0640$  & $-26.7$ & $-0.42$ \\
 2234$+$282 & $-0.0461$ & $-0.0218$ & $2.85$  & $-0.60$ \\
 0613$+$570 & $0.0528$  & $0.0349$  & $-65.9$ & $0.22$  \\
 0529$+$483 & $0.0127$  & $-0.0095$ & $61.7$  & $0.93$  \\
 0642$+$449 & $0.0126$  & $0.0352$  & $-11.9$ & $0.37$  \\
\hline
\end{tabular}
\caption{Source properties used in our VGOS simulation. From left to right, source name (which includes its coordinates), fractional Stokes Q, fractional Stokes U (both at 4\,GHz), Rotation Measure and spectral index.}
\label{VGOSTab}
\end{table}

Obviously, the SNR of the fringes is too low for an independent fit of the D-terms to each spectral window and source (which is the kind of fit that 
would be performed with the classical \texttt{LPCAL} software). However, we used \texttt{polsolve} to fit the whole dataset (i.e., all sources and spectral windows in one simultaneous fit) by imposing consistent source polarization along the whole observing bandwidth (i.e., the same polarization parameters, including frequency de-polarization and/or Faraday rotation, across the band). This way, the effective SNR used in the fit corresponds to the whole bandwidth (i.e., SNR $\sim 50$ per scan). For each source, we model the Stokes parameters at a fiducial frequency (equal to the lowest frequency in the data), together with a Rotation Measure (i.e., $RM = \frac{\partial \chi}{\partial \lambda^2}$) and a polarization spectral index, $\alpha$ (i.e., $p \propto \nu^\alpha$), which will account for frequency (de)polarization. These quantities allow us to model the source polarization in the whole observing band using only four parameters. Regarding the D-terms, we allow for independent values at each frequency channel (i.e., we use the ``multi-IF'' fitting approach mentioned in Sect. \ref{PSolveDescSec}). 

\begin{figure*}
\centering
\includegraphics[width=0.99\textwidth]{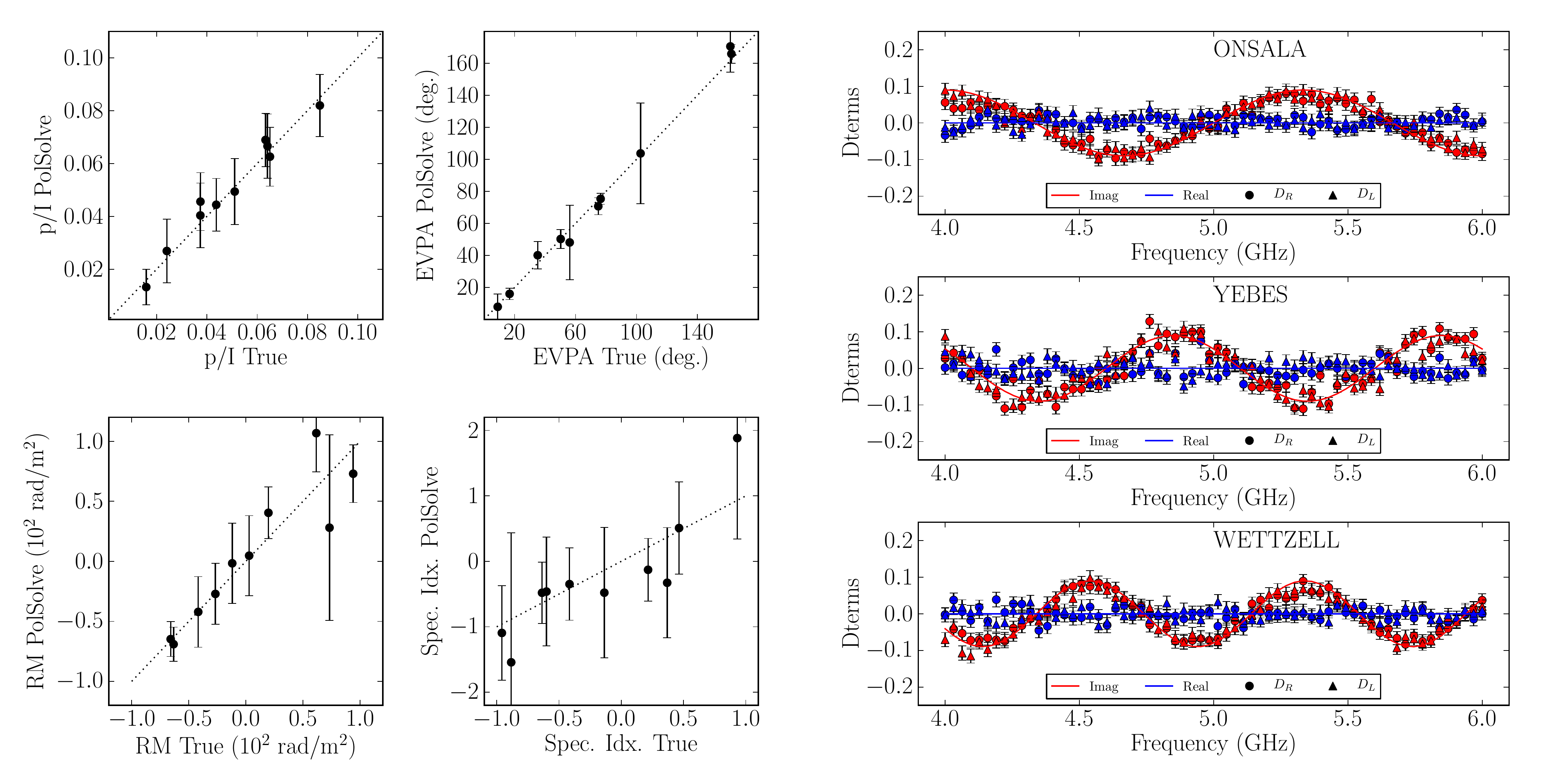}
\caption{ Results of the \texttt{polsolve} fit to the simulated VGOS observations. The left panels show correlation plots between the true source polarization quantities (fractional polarization and EVPA at the fiducial frequency, Rotation Measure and polarization spectral index) and the \texttt{polsolve} estimates. Dotted lines mark the 1:1 correlation. The right panels show the simulated D-term spectra as thick lines (resulting from a frequency-dependent phase offset between the linear polarizers of the antennas) and the values fitted by \texttt{polsolve}.}
\label{VGOSFig}
\end{figure*}

The result of the global fit with \texttt{polsolve} is shown in Fig. \ref{VGOSFig}. We notice that the model to the whole dataset consists of 768 parameters for the instrumental polarization (i.e., two complex numbers for each antenna and frequency channel) plus 40 parameters for the source polarization (i.e., four quantities per source, which are two Stokes parameters, Q and U, a Rotation Measure and a polarization spectral index). This number of parameters is much lower than the total of 7700 parameters that would be needed by \texttt{LPCAL} (i.e., one independent fit for each source and spectral channel). In the left panels of Fig. \ref{VGOSFig}, we show correlation plots between the true source polarization quantities and the values fitted with \texttt{polsolve}. The error bars shown in Fig. \ref{VGOSFig} are derived by \texttt{polsolve} from the postfit covariance matrix, by assuming a reduced $\chi^2$ equal to its expected value \cite[similar to the strategy followed in ][]{UVMultiFit}. In the figure panels at right, we show the simulated D-terms of all three antennas, together with the values fitted with \texttt{polsolve}.  It is remarkable that the program is able to recover the source polarization parameters and the instrumental polarization using only two scans per source (and with such a low SNR per spectral window). The use of several calibrators in one simultaneous fit allows us to maximize the information in the fit, in such a way that, even with only two scans per source, we are able to decouple the effects of the instrumental polarization from the source contribution in the cross-hand visibilities.

We note that in the cases of wider bandwidths, which could even cover different radio bands, any contribution of internal Faraday rotation in the calibrators may cause their EVPAs to depart from the canonical $\lambda^2$ law, hence biasing the \texttt{polsolve} model. In addition, opacity effects in the jets may introduce position shifts of the brightness peaks as a function of frequency, which would bias the D-term estimates if not taken into account. In these cases, the use of polarization self-calibration (see Sect. \ref{PolSelfCal}), based either on a (full-polarization) multi-frequency-synthesis deconvolution \citep{MFSRef} or on the construction of an image ``cube'' (i.e., different model images per band), would be an alternative way to calibrate the VGOS instrumental polarization with \texttt{polsolve}, which would still allow for the use of the ``multi-IF'' mode.

\subsection{Complex Source Structures: EHT Simulations}

The Event Horizon Telescope (EHT) is an Earth-size mm-VLBI array,
whose main goal is the imaging of horizon-scale structures in close by super-massive black holes \citep{EHT1}. In this paper, we simulated EHT observations following the array configuration of the 2017 campaign, consisting of eight observatories at six geographical locations \citep{EHT2}: Atacama Large Millimeter/sub-millimeter Array (ALMA) and the Atacama Pathfinder Experiment (APEX) in the Atacama Desert in Chile; the Large Millimeter Telescope Alfonso Serrano (LM) on the Volc\'an Sierra Negra in Mexico; the Arizona Radio Observatory (AZ), situated on Kitt Peak, Arizona; the South Pole Telescope (SPT) at the Geographic South Pole; the James Clerk Maxwell Telescope (JCMT) and Sub-millimeter Array (SMA) on Maunakea in Hawai'i (USA) and the IRAM 30m telescope, on Pico Veleta in Spain. Their respective antenna mounts (which affect the feed angles and, as a consequence, the polarimetry calibration) are given in Table~\ref{EHTantennas}. All these antennas (with the exception of ALMA) used circular polarization feeds. The visibilities related to ALMA were converted into a pure circular basis using the \texttt{polconvert} program \citep{POLCONVERTREF,APPRef}, which may introduce a small pure-imaginary and symmetric instrumental polarization at ALMA \citep[e.g., ][]{CiriacoRef}.

\begin{table}
\centering
\begin{tabular}{l|l}
 Telescope & Mount\\ \hline 
 ALMA & Alt-Azimuth\\
 APEX & Nasmyth-Right\\
 SMT & Nasmyth-Right\\
 JCMT & Alt-Azimuth\\
 LMT & Nasmyth-Left\\
 SMA & Nasmyth-Right\\
 SPT & Alt-Azimuth\\
 IRAM-30m & Nasmyth-Left\\ \hline 
\end{tabular}
\caption{Names of the EHT telescopes that participated in the 2017 campaign and their respective antenna mounts.}
\label{EHTantennas}
\end{table}

\subsubsection{Generating synthetic EHT data}

We used the \texttt{polsimulate} task to generate synthetic EHT observations of a polarized source with a complex structure, in order to test the limitations of the \texttt{LPCAL} algorithm for the parametrization of the source polarization brightness distribution (i.e., the source partition approach formulated in Eq. \ref{PSolveIEq} and Eq. \ref{PSolveQUEq}) and compare its performance to the iterative polarimetry self-calibration approach. 

The whole set of antennas that participated in the EHT 2017 observing campaign were included in the simulations.
There is only one source in these synthetic data, with its coordinates set to those of the Galactic Center (Sagittarius A$^*$). The exact observing times are set to those of the real EHT observations of the Galactic Center taken on April 7, 2017. The observing frequency is set to 230\,GHz, with a bandwidth of 2\,GHz, and the system temperatures, $T_{sys}$, of all antennas are set to a single value of 50\,K. Although the EHT is a heterogeneous array, with different ranges of $T_{sys}$ at each station, and a $T_{sys}$ of 50\,K is a rather low value for a receiver at 230\,GHz \citep[only the Phased ALMA approaches that value; see, e.g. Table 2 of][]{EHT3}, we notice that real calibrator observations with the EHT \citep[e.g., 3C\,279;][]{EHT3C279} are currently limited by dynamic range (i.e., sensitivity may  not be the main limitation in the calibration of the instrumental polarization of the EHT). 

The visibility noise is estimated by \texttt{polsimulate} using the CASA simulation tool, \texttt{sm}, based on the effective diameters of the telescopes \citep{EHT2} and the $T_{sys}$. The real and imaginary parts of the D-terms for all the antennas are taken from a random Gaussian distribution, centered at zero and with a standard deviation of 0.2. We notice that the Dterms resulting from this distribution are usually large, compared to the typical D-terms of EHT antennas \citep[e.g.][]{Johnson15, EHT7} and are not representative of the true instrumental polarization of the EHT.

The simulated source structure has a core-jet shape oriented in the East-West direction. There are three unpolarized point-like components (a 1\,Jy core plus two 0.8\,Jy jet components, separated by 40 and 80 $\mu$as from the core) and two polarized components (at 10 and 30 $\mu$as from the core, with 0.4\,Jy each) with high fractional polarizations (0.94 and 0.90) and different EVPAs (16\,deg. and $-45$\,deg.). In Fig. \ref{fig:ehtmap}, we show the CLEAN reconstruction (using uniform visibility weighting) of the simulated data after correcting for the effects of the instrumental polarization. The distribution of the source components has been made in such a way that the peaks of polarized intensity (marked with yellow crosses in the figure) are shifted with respect to the peaks in total intensity. These shifts in polarization brightness are a problem for the \texttt{LPCAL} source parametrization, since the polarization structure is not proportional to that of Stokes $\mathcal{I}$, either across the extent of the core or on each of the jet components. Therefore, the D-term estimates resulting from the \texttt{LPCAL} approach may be biased by the contribution of the source structure that, due to the shifts in the polarization intensity, cannot be reproduced by the source model (Eqs. \ref{PSolveQUEq}).

\subsubsection{Polarimetry self-calibration}
\label{PolSelfCal}

\begin{figure}
    \centering
        \includegraphics[width=0.49\textwidth]{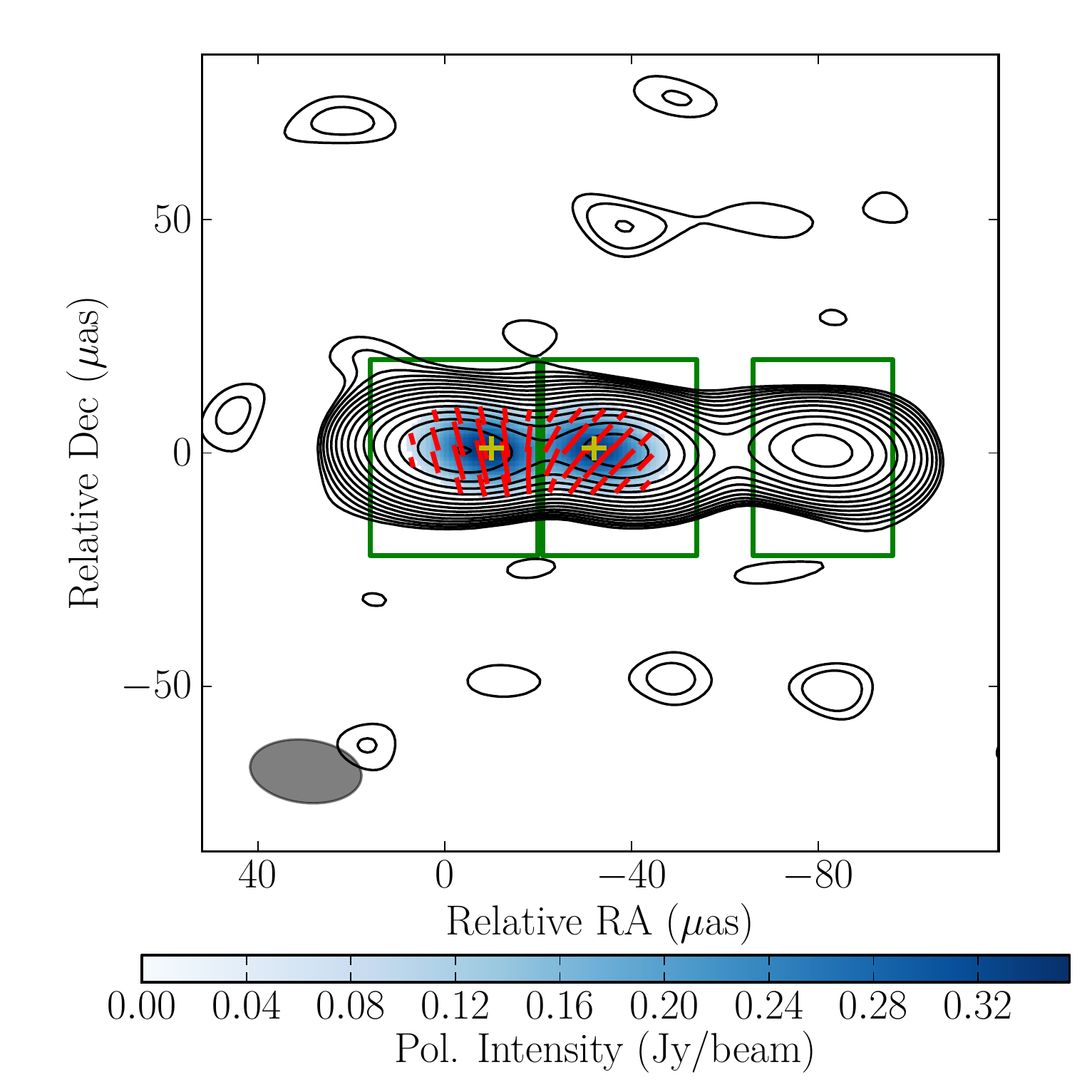}
    \caption{CLEAN image of the synthetic EHT observations free from instrumental polarization. The FWHM of the restoring beam is shown at bottom-left. The 16 contour levels are separated in logarithmic scale and represent total intensity (peak of 1.28\,Jy/beam; lowest contour of 1.3\,mJy/beam); blue color scale represents polarized intensity (peak of 0.36\,Jy/beam) and red lines represent the EVPA spatial distribution. The yellow crosses indicate the locations of the polarization intensity peaks, which are shifted with respect to those of Stokes $\mathcal{I}$. The green boxes indicate the extent of the polarization sub-components used in the D-term fitting with the \texttt{LPCAL} approach.}
    \label{fig:ehtmap}
\end{figure}

\begin{figure*}
    \centering
        \includegraphics[width=\textwidth]{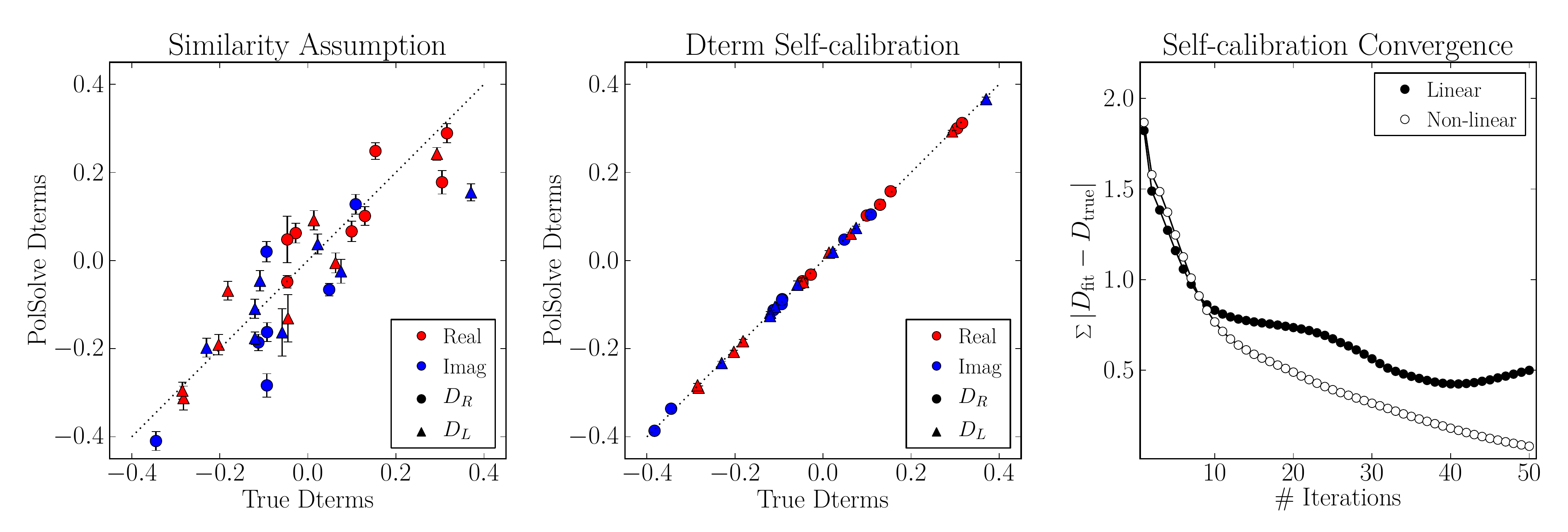}
    \caption{Results of \texttt{polsolve} applied to our simulated EHT observations. Left, correlation between the true antenna D-terms and the values fitted using \texttt{LPCAL}'s similarity assumption (see text). Center, correlation for the D-terms fitted after fifty iterations of polarimetry self-calibration. Right, $L_1$ norm between true and fitted D-terms, as a function of the number of self-calibration iterations, for the case of linear D-term approximation (black points) and full non-linear D-term model (white points).}
    \label{fig:selfcal}
\end{figure*}

For the D-term fitting using the \texttt{LPCAL} approach, we have divided the source structure into three polarization sub-components. These sub-components are shown as green rectangles in Fig. \ref{fig:ehtmap}. Each rectangle is centered on each of the main total-intensity peaks of the source. The results of the D-term fitting using this source parametrization are shown in Fig. \ref{fig:selfcal} (left panel). Even though there is some correlation between the true and the estimated D-terms, the deviations in the correlation (which, in some cases, are as high as 20\%) may introduce high residual instrumental effects in the cross-hand visibilities, which could limit the dynamic range (and fidelity) of the polarization image. The quality of the fit could be improved by increasing the number of polarization sub-components (i.e., dividing the green rectangles of Fig. \ref{fig:ehtmap} into smaller pieces), but the final result would still depend on the exact location and extent of the sub-components with respect to the location of the true polarization features of the source. In any case, the simple source subdivision used in this example reflects the limitations of the \texttt{LPCAL} strategy in a clear way. 

We can quantify the quality of the D-term fitting using the $L_1$ norm of the D-term residuals (the difference between true and estimated D-terms), i.e.

\begin{equation}
L_1 = \sum_{\mathrm{ants}}{\left| D_{\mathrm{fit}} - D_{\mathrm{true}}    \right|}.
\label{L1Eq}
\end{equation}

\noindent Using the D-term estimates from the \texttt{LPCAL} approach on our EHT simulated data (Fig. \ref{fig:selfcal}, left), the $L_1$ norm obtained is $L_1 = 1.87$.

In contrast to these results, we show in Fig. \ref{fig:selfcal} (center panel) the correlation between true D-terms and the \texttt{polsolve} D-term estimates obtained after fifty iterations of the polarimetry self-calibration approach (using the CLEAN algorithm as implemented in CASA). The correlation is, in this case, remarkably better than that
obtained with the \texttt{LPCAL} calibration approach. 
We can reach values as low as $L_1 = 0.2$ (i.e., an order of magnitude lower than with the \texttt{LPCAL} approach) after 20$-$30 iterations.

\subsubsection{Linear vs. non-linear leakage Determinations}
\label{SimulNLSec}

The bias in the D-term estimates produced by the \texttt{LPCAL}'s linear approximation of Eq. \ref{Chi2PsolveEq} may become important for cases of antennas with relatively high polarization leakage (D-terms higher than 10$-$20\%) and/or sources with high fractional polarizations in Fourier space \citep[a quantity that, in principle, is unbounded, because of the possible presence of low visibility values, or even visibility nulls, in the parallel-hand correlations; ][]{Johnson15}.

The effects of the non-linear contribution of the D-terms in the cross-hand visibilities of a VLBI dataset can be quantified with \texttt{polsolve}. It is possible to disconnect the non-linear D-term contributions from the fit (this behavior is controlled by a special keyword in the CASA task), which allows us to compare directly the $L_1$ norm of the D-term residuals with and without non-linear corrections. In Fig. \ref{fig:selfcal} (right panel), we show the $L_1$ norm of the D-term residuals (true values minus fitted values), as a function of the number of polarization self-calibration iterations, for the fitting with linear approximation of the D-term effects (black points) and for the fitting with the full \texttt{polsolve} D-term model (white points).

For a small number of iterations, both models give D-term estimates of similar quality. However, as the number of iterations increases, it becomes clear that the limitations of the D-term linear approximation do not produce any improvement of the calibration of the cross-hand data beyond a given limit (i.e., there is a maximum achievable dynamic range in the model image, which only appears in the case of the D-term linear approximation). On the other hand, if the full D-term model is used in the fit, the full-polarization image of the source keeps improving with the number of self-calibration iterations (and, as a consequence, the estimated D-terms keep approaching the true simulated values), with no appreciable saturation in the first fifty self-calibration iterations. We notice, though, that other factors (e.~g., an imperfect phase self-calibration) may contribute to increase the $L_1$ norm in real observations.

A direct conclusion from these results is that the non-linear D-term model of \texttt{polsolve} allows us to reach higher dynamic ranges in the polarization images obtained from high-sensitivity VLBI observations, assuming that the antenna gains are properly calibrated.

\section{Real Observations}

In this section, we present the results of \texttt{polsolve} applied to real VLBI experiments, for which full-polarization images are available in the literature. Since the published images are obtained from visibilities calibrated with \texttt{LPCAL}, we can test the performance of \texttt{polsolve} by comparing them to the results obtained in CASA.

\subsection{3C\,345 at 86\,GHz} 
\label{sec:GMVA}

The Global mm-wave VLBI Array (GMVA) is an international collaboration led by the Max-Planck-Institut f\"ur Radioastronomie (MPIfR)\footnote{See \texttt{http://www.mpifr-bonn.mpg.de/div/vlbi/globalmm}.}, between NRAO and several European institutions, aimed at providing an infrastructure for routine VLBI observations at millimeter wavelengths. In MV12, a full-polarization calibration pipeline was presented for the GMVA. That pipeline was based on AIPS and made use of the \texttt{LPCAL} program for the determination of the antenna D-terms. One of the conclusions from MV12 is that, even though there were some inconsistencies in the D-terms determined from different sources observed within the same experiment, the final images of sources with a high linear-polarization brightness did not depend strongly on the actual combination of D-term values used in the calibration. A rule of thumb from the results presented in MV12 is that any polarization intensity higher than $\sim$50\% of the polarization peak has a very weak dependence on the different GMVA D-term estimates.

\begin{figure*}
    \centering
        \includegraphics[width=0.4\textwidth]{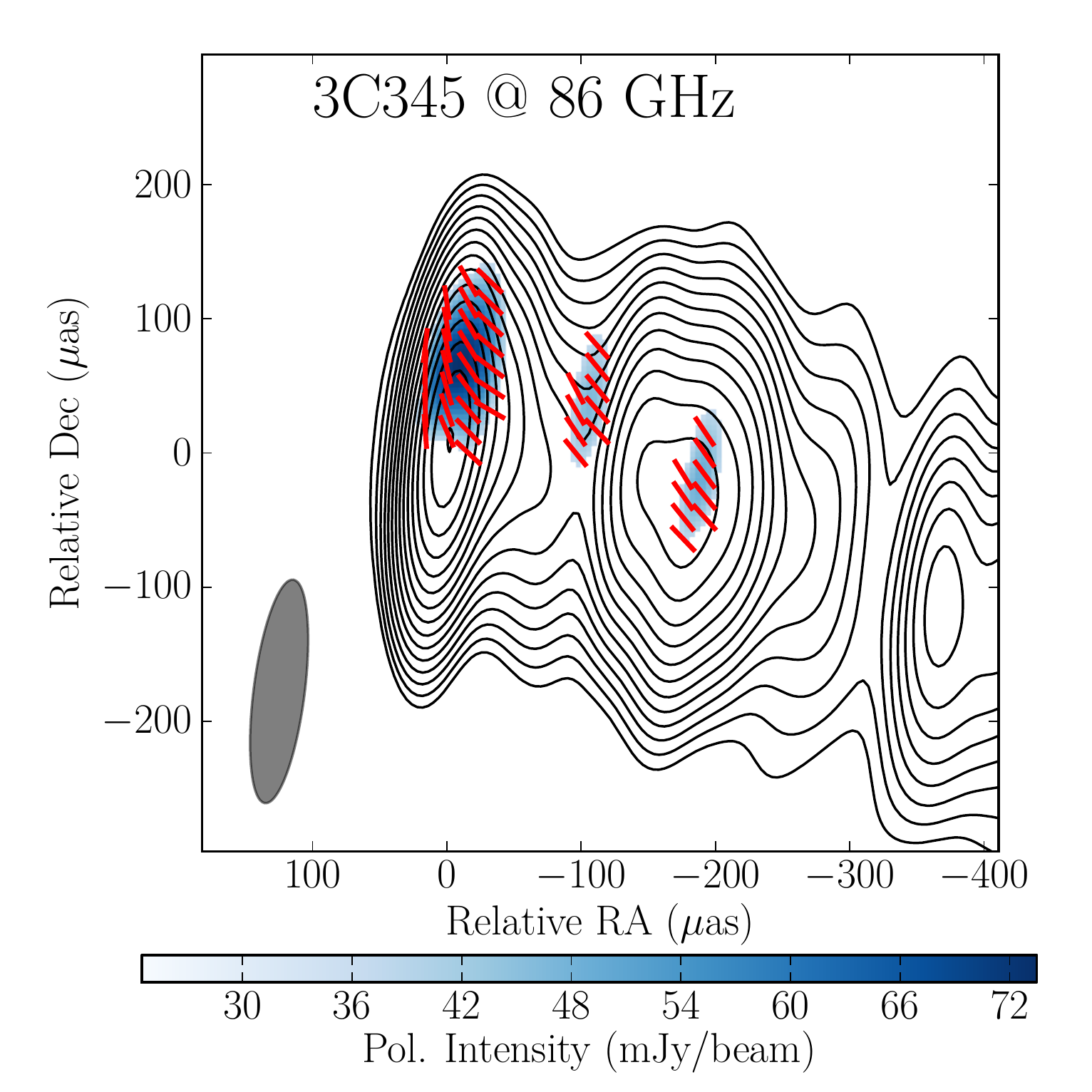}
        \includegraphics[width=0.4\textwidth]{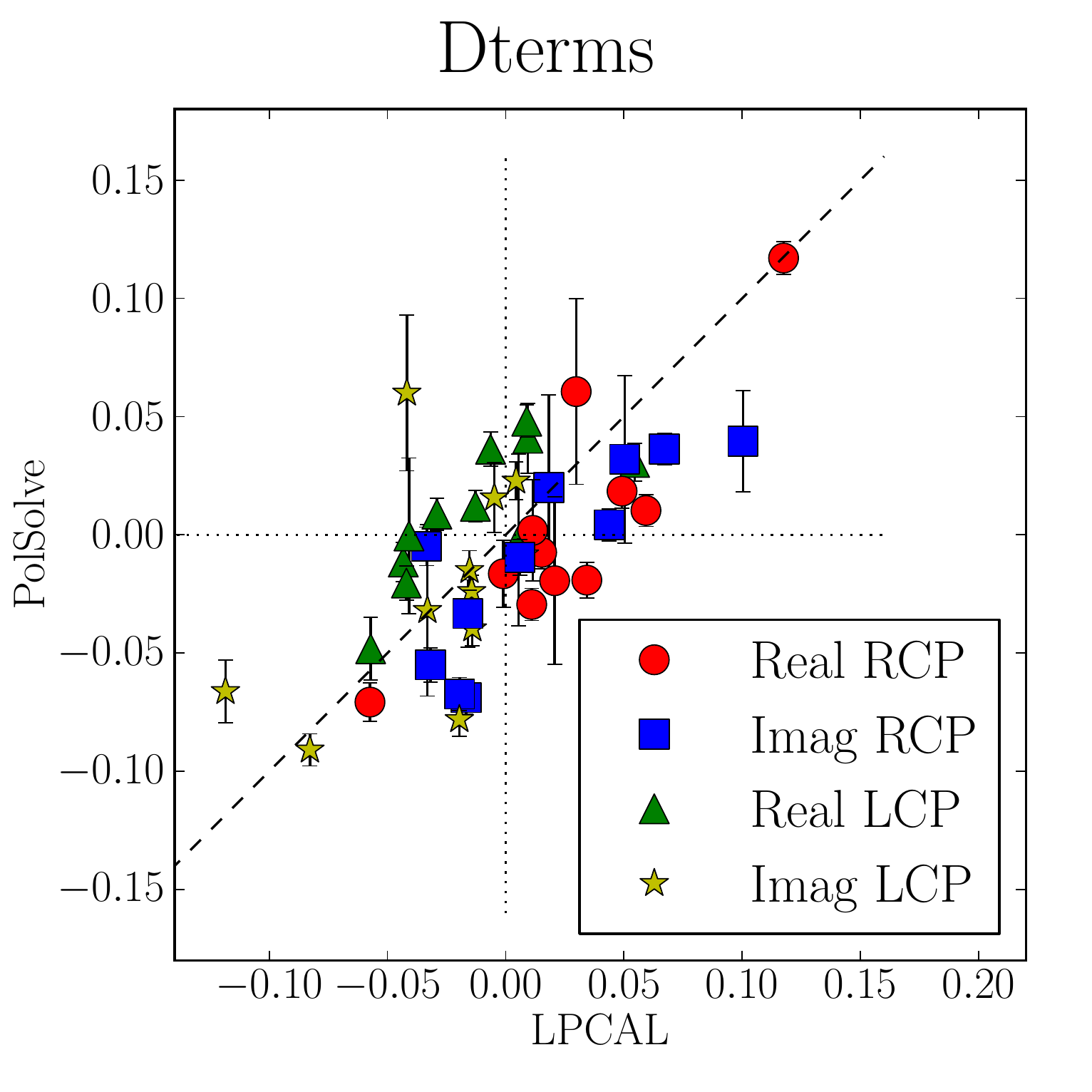}
    \caption{Left, GMVA image of 3C\,345 obtained after the \texttt{polsolve} calibration of the observations reported in MV12. FWHM of convolving beam is shown at bottom-left. Contours are logarithmically spaced between 35\,mJy/beam and 1.22\,Jy/beam (the peak intensity).  Right, correlation between the \texttt{LPCAL} and \texttt{polsolve} estimated D-terms.}
    \label{fig:GMVA}
\end{figure*}

We have reprocessed the data reported in MV12, which corresponds to the GMVA campaign of May 2010. We refer to that publication for the details of these observations (i.e., participating antennas, frequency configuration, scheduling issues, etc.). We have imported the fringe-fitted visibilities of source 3C\,345 (for which a full-polarization image is shown in Fig. 9 of MV12) 
into CASA format, to perform the hybrid imaging and polarimetry calibration. We have re-imaged and self-calibrated the visibilities using the CASA tasks \texttt{tclean}, \texttt{gaincal} and \texttt{applycal}, by combining the two parallel-hand correlations for the gain solutions (to not affect the instrumental polarization) and using Briggs weighting \citep{BRIGGS} with a robustness parameter of 0.5 (which gives a good trade-off between sensitivity and spatial resolution). Then, we have executed a set of twenty iterations of polarimetry self-calibration (see Sect. \ref{PolSelfCal}), which resulted in a fast and robust convergence of the antenna D-terms (the estimated values did not change above numerical noise after just $\sim 5$ iterations). The resulting image and D-terms are shown in Fig. \ref{fig:GMVA}. We have applied a polarization cutoff of 50\% of the peak, the same as in Fig. 9 of MV12. Even though the synthesized CLEAN beam of our image is different from that of Fig. 9 in MV12, the main features of the source (in total intensity and in linear polarization) are similar between the two images. In particular, there is linear polarization at the north of the core (and a bit upstream of the jet) with a clear bending of the EVPA, from parallel to the jet to perpendicular. Then, at about 100 and 200\,$\mu$as downstream from the core, another two polarized components appear, with their EVPAs oriented parallel to the jet direction.

Regarding the D-terms, we have executed \texttt{LPCAL} by dividing the core region of 3C\,345 into two subcomponents, one that covers the northern side of the jet (where there is the maximum polarization brightness) and the other one that covers the southern part (where there is no detected polarization). We have also set other subcomponents, which cover the two polarization features downstream from the core. We notice that it is not possible to model the EVPA rotation in the core with such a source subdivision, so we expect to have a degradation in the quality of the D-term estimates with \texttt{LPCAL}. The D-term estimates are shown in Fig. \ref{fig:GMVA} (right). Similar to the case discussed in Sect. \ref{PolSelfCal} (see Fig. \ref{fig:selfcal}, left), the correlation between the D-terms estimated with the \texttt{LPCAL} approach and those from the polarization self-calibration is rather poor (Fig. \ref{fig:GMVA}), likely due to the limited accuracy of the former method to account for the source polarization structure. In any case, the resemblance between the image obtained with AIPS/\texttt{lpcal} (shown in MV12) 
and the image obtained with CASA/\texttt{polsolve} (Fig. \ref{fig:GMVA}) indicates a satisfactory calibration of the GMVA data with \texttt{polsolve}.

\subsection{3C\,279 at 43\,GHz} 
\label{sec:VLBA}

K20 report on several VLBA full-polarization observations at wavelengths of 7\,mm and 1.3\,cm of the active nucleus in M87. We have downloaded the data corresponding to their project BG251A from the NRAO archive\footnote{\texttt{https://science.nrao.edu/observing/data-archive}}. This project was observed in May 2017 at 7\,mm (i.e., 43\,GHz) and included several scans of source 3C\,279, used as a polarization calibrator. The frequency configuration was arranged in eight contiguous subbands of equal width, covering a total of 256\,MHz (see K20 for more details on the observations).

\begin{figure*}[th!]
    \centering
        \includegraphics[width=0.4\textwidth]{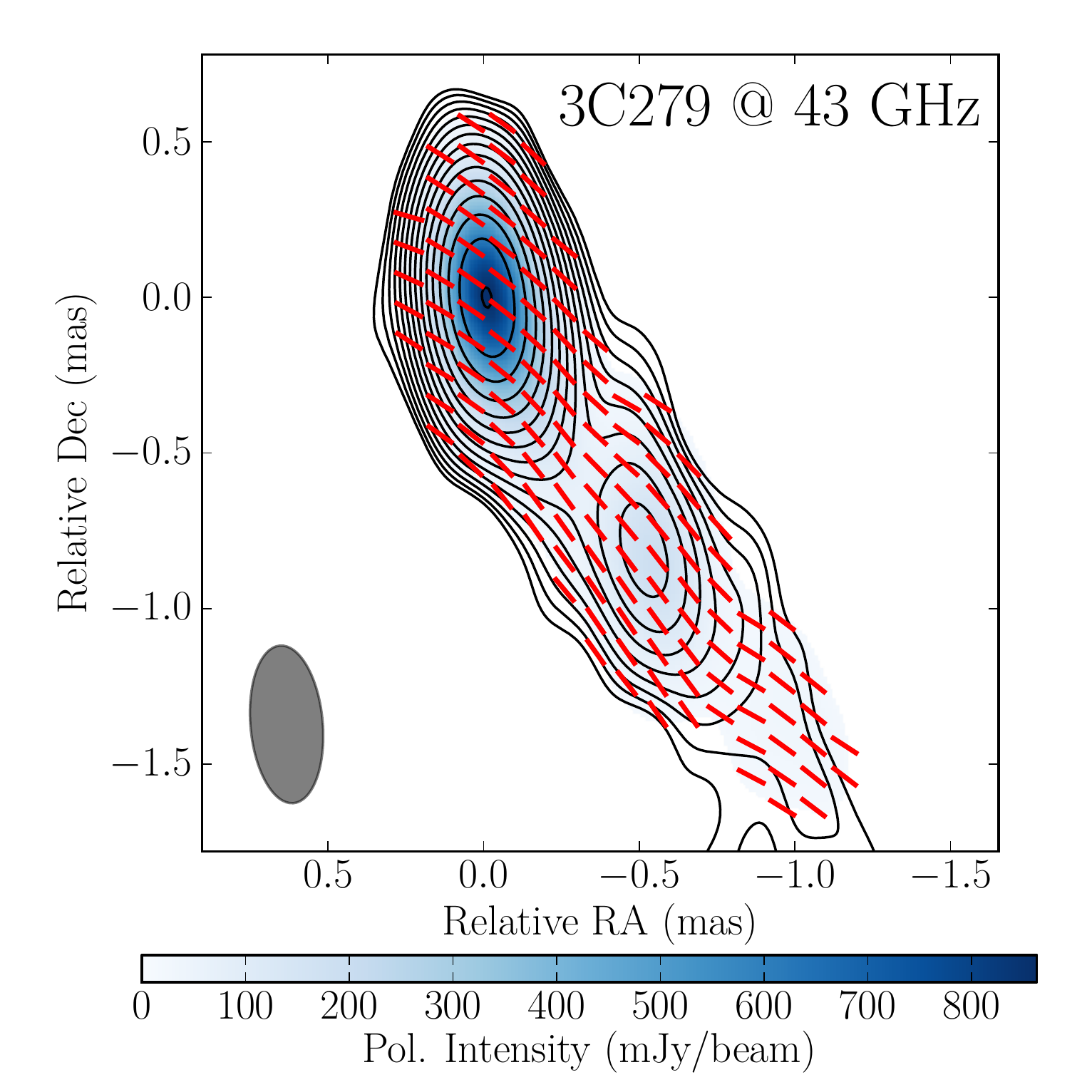}
        \includegraphics[width=0.4\textwidth]{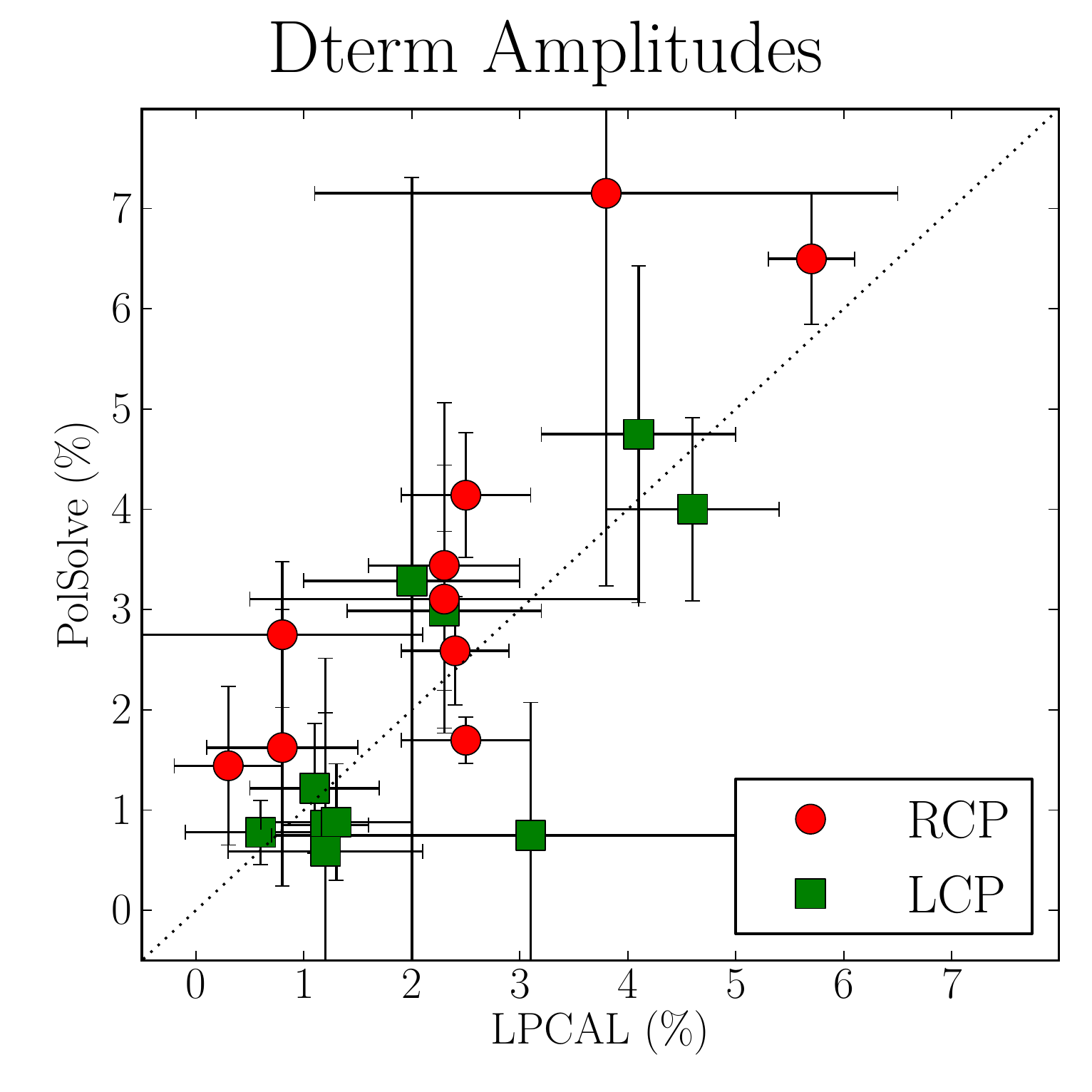}
    \caption{Left, VLBA image of 3C\,279 obtained from the \texttt{polsolve} calibration of the observations reported in K20. FWHM of convolving beam is shown at bottom left. Contours are logarithmically spaced between 50\,mJy/beam and 8.4\,Jy/beam (the peak intensity). Right, correlation between the D-term amplitudes reported in K20 (see their Table A.2) and the amplitudes of the frequency-averaged ``multi-IF'' D-terms estimated with \texttt{polsolve}.}
    \label{fig:VLBA}
\end{figure*}

We have reprocessed the BG251A observations of source 3C\,279, in order to compare the \texttt{polsolve} D-term calibration to that obtained with \texttt{LPCAL} (as used by K20). 
The amplitude calibration, global fringe-fitting and cross-polarization delay corrections were done in AIPS using standard procedures. Then, the data were exported into CASA (version 5.7) for hybrid imaging (several iterations of phase and amplitude self-calibration), using also standard procedures and Briggs weighting with a robustness parameter of 0.5. After the imaging, we executed 10 iterations of polarization self-calibration with \texttt{polsolve}, using the ``multi-IF'' mode. The resulting CLEAN image in \texttt{mfs} mode, shown in Fig. \ref{fig:VLBA} (left), is remarkably similar to the results reported in K20 (see their Fig. A.2, lower-left panel). Regarding the D-term estimates (Fig. \ref{fig:VLBA}, right), the \texttt{polsolve} values are correlated with those reported in K20 (see their Table A.2), although the correlation is also rather poor. The worst outliers in the correlation happen to be the cases with the largest D-term dispersion among subbands (i.e., the cases with the largest error bars, see Table A.2 of K20), which correspond to the antennas at Hancock (RCP polarizer) and St. Croix (LCP polarizer).

\section{Summary and Conclusions}
\label{sec:summary}

New VLBI arrays, with high sensitivities and wide instantaneous bandwidths, as compared to the state-of-the-art in the past decades, require the use of improved calibration algorithms to account for the effects of instrumental polarization. In the case of wide bandwidths and geodesy-like observations (where many weak sources are observed with a limited parallactic-angle coverage), an algorithm able to combine the wide bandwidths (to increase the solution SNR) and the use of several calibrators simultaneously (to maximize the amount of information in the fit) is required. In the case of sensitive arrays and/or interferometers with very high spatial resolutions, special algorithms that account for the source polarization structures and correct for the non-linear effects of the instrumental polarization may also be needed.

We present a new set of algorithms, implemented for use in the CASA interface and fully available to the VLBI community, that allow the calibration of instrumental polarization in VLBI observations to be performed without the limitations of the classical calibration algorithms used so far. Our algorithms (a realistic simulator, \texttt{polsimulate}, a flexible solver, \texttt{polsolve}, and several other helper functions) allow us to perform advanced polarization calibration free of the limiting assumptions of the standard calibration software (mainly, observations with narrow fractional bandwidths, no use of a-priori information and limited sensitivity in the polarization signal). We show the results of our software applied to realistic VLBI simulations and compare its performance with that of the most used VLBI polarimetry algorithm, \texttt{LPCAL}. We also apply \texttt{polsolve} to real mm-wave VLBI observations of sources 3C\,345 and 3C\,279, and obtain full-polarization images similar to those already published (MV12, K20). 

The advantages of \texttt{polsolve} may be critical for the optimum analysis of polarization observations taken with new-generation VLBI arrays.

\begin{acknowledgement}
This work has been partially supported by the MICINN Research Project PID2019-108995GB-C22. IMV and AM also thank the Generalitat Valenciana for funding, in the frame of the GenT Project CIDEGENT/2018/021. MJ is supported by the ERC Synergy Grant ``BlackHoleCam: Imaging the Event Horizon of Black Holes'' (Grant 610058).
\end{acknowledgement}


\begin{thebibliography}{}

\bibitem[Alef et al.(2019)]{AlefEUVGOS}{Alef, W., Anderson, J.~M., Bernhart, S., et al.\ 2019, Proceedings of the 24th European VLBI Group for Geodesy and Astrometry Working Meeting, 107}

\bibitem[van Bemmel et al.(2018)]{vanBemmel2018}{van Bemmel, I., Small, D., Kettenis, M., et al.\ 2018, 14th European VLBI Network Symposium \& Users Meeting (EVN 2018), 79}

\bibitem[Briggs(1995)]{BRIGGS} Briggs, D. S. 1995, Ph. D. thesis, New Mexico Institute of Mining and Technology


\bibitem[Broderick et al.(2020)]{ThemisRef} Broderick, A.~E., Gold, R., Karami, M., et al.\ 2020, \apj, 897, 139


\bibitem[Chael et al.(2018)]{EHTimRef}{Chael, A., Bouman, K., Johnson, M. et al.\ 2018, Eht-imaging: Tools for Imaging and Simulating VLBI Data, 1.0, Zenodo, doi:https://doi.org/10.5281/zenodo.1173414}


\bibitem[Cotton(1993)]{Cotton93}{Cotton, W.~D.\ 1993, \aj, 106, 1241}

\bibitem[EHT Collaboration et al.(2019a)]{EHT1}{ Event Horizon Telescope Collaboration et al.\ 2019, \apjl, 875, L1}

\bibitem[EHT Collaboration et al.(2019b)]{EHT2}{ Event Horizon Telescope Collaboration et al.\ 2019, \apjl, 875, L2}

\bibitem[EHT Collaboration et al.(2019c)]{EHT3}{ Event Horizon Telescope Collaboration et al.\ 2019, \apjl, 875, L3}

%\bibitem[EHT Collaboration et al.(2019d)]{EHT4}{ Event Horizon Telescope Collaboration et al.\ 2019, \apjl, 875, L4}

%\bibitem[EHT Collaboration et al.(2019e)]{EHT5}{ Event Horizon Telescope Collaboration et al.\ 2019, \apjl, 875, L5}

%\bibitem[EHT Collaboration et al.(2019f)]{EHT6}{ Event Horizon Telescope Collaboration et al.\ 2019, \apjl, 875, L6}

\bibitem[EHT Collaboration et al.(2020a)]{EHT3C279}{ Event Horizon Telescope Collaboration et al.\ 2020, \aap, 640, A69}

\bibitem[EHT Collaboration et al.(2020b)]{EHT7}{ Event Horizon Telescope Collaboration et al.\ 2020, in preparation}

\bibitem[Fey et al.(2015)]{ICRFREF}{Fey, A.~L., Gordon, D., Jacobs, C.~S., et al.\ 2015, \aj, 150, 58}

\bibitem[Goddi et al.(2019)]{CiriacoRef}{Goddi, C., Mart{\'\i}-Vidal, I., Messias, H., et al.\ 2019, \pasp, 131, 075003}

\bibitem[G{\'o}mez et al.(2016)]{GomezRef} G{\'o}mez, J.~L., Lobanov, A.~P., Bruni, G., et al.\ 2016, \apj, 817, 96

\bibitem[Hamaker et al.(1996)]{Hamaker} Hamaker, J.~P., Bregman, J.~D., \& Sault, R.~J.\ 1996, \aaps, 117, 137



\bibitem[H{\"o}gbom(1974)]{Hogbom}{H{\"o}gbom, J.~A.\ 1974, \aaps, 15, 417}

\bibitem[Janssen et al.(2019)]{Janssen2019}{Janssen, M., Goddi, C., van Bemmel, I. et al.\ 2019, \aap, 626, A75}

\bibitem[Johnson et al.(2015)]{Johnson15}{Johnson, M.~D., Fish, V.~L., Doeleman, S.~S., et al.\ 2015, Science, 350, 1242
}

\bibitem[Kravchenko et al.(2020)]{KravVLBA} Kravchenko, E., Giroletti, M., Hada, K. et al. 2020, \aap,  637, L6

\bibitem[Leppanen et al.(1995)]{Leppanen95}{Leppanen, K.~J., Zensus, J.~A., \& Diamond, P.~J.\ 1995, \aj, 110, 2479}

\bibitem[Mart{\'\i}-Vidal et al.(2012)]{IMVGMVA}{Mart{\'\i}-Vidal, I., Krichbaum, T.~P., Marscher, A., et al.\ 2012, \aap, 542, A107}

\bibitem[Mart{\'\i}-Vidal et al.(2014)]{UVMultiFit}{Mart{\'\i}-Vidal, I., Vlemmings, W.~H.~T., Muller, S., et al.\ 2014, \aap, 563, A136}

\bibitem[Mart{\'\i}-Vidal et al.(2016)]{POLCONVERTREF}{Mart{\'\i}-Vidal, I., Roy, A., Conway, J., et al.\ 2016, \aap, 587, A143}

\bibitem[Matthews et al.(2018)]{APPRef}{Matthews, L.~D., Crew, G.~B., Doeleman, S.~S., et al.\ 2018, \pasp, 130, 015002}

\bibitem[Molina et al.(2014)]{AgudoGMVA}{Molina, S.~N., Agudo, I., G{\'o}mez, J.~L., et al.\ 2014, \aap, 566, A26}

\bibitem[Park et al.(2020)]{Jongho}{Park, J. et al.\ 2020, \apj (submitted)}

\bibitem[Pearson \& Readhead(1984)]{selfcalRef}{Pearson, T.~J. \& Readhead, A.~C.~S.\ 1984, \araa, 22, 97}

\bibitem[Pesce(2020)]{dom2020} Pesce, D.\ 2020, in preparation

\bibitem[Petrachenko et al.(2012)]{VGOSREF}{Petrachenko W. T., Niell A. E., Corey B. E., et al. 2012, Proc. of Geodesy for Planet Earth, IAGS, 136, 2012}

\bibitem[Rau \& Cornwell(2011)]{MFSRef} Rau, U. \& Cornwell, T.~J.\ 2011, \aap, 532, A71


\bibitem[Smirnov(2011)]{Smirnov}{Smirnov, O.~M.\ 2011, \aap, 527, A106}

\bibitem[O'Sullivan et al.(2011)]{Sullivan11} O'Sullivan, S.~P., Gabuzda, D.~C., \& Gurvits, L.~I.\ 2011, \mnras, 415, 3049


\bibitem[Thompson et al.(2017)]{TMS}{Thompson, A.~R., Moran, J.~M., \& Swenson, G.~W.\ 2017, Interferometry and Synthesis in Radio Astronomy, by A. Richard Thompson, James M. Moran, and George W. Swenson, Jr. 3rd ed. Springer, 2017}






\end{thebibliography}
\end{document}